\documentclass[11pt]{article}
\usepackage{moriond,epsfig}

\def\ra{\rightarrow}

\def\mus{$\mu$s}

\def\Do{D^0}

\def\be{\begin{equation}}
\def\ee{\end{equation}}
\def\bea{\begin{eqnarray}}
\def\eea{\end{eqnarray}}

\begin{document}
\vspace*{4cm}
\title{THE START OF RUN II at CDF}

\author{ M. Rescigno }

\address{I.N.F.N. Sezione di Roma and Fermilab,\\
PO Box 500 Batavia IL, 60510, USA}

\maketitle\abstracts{
After a hiatus of almost 6 years and an extensive upgrade, Tevatron, 
the world largest proton--antiproton collider, 
has resumed the operation for the so called RUN II. 
In this paper we give a brief overview  of the many new 
features of the Tevatron complex and of the upgraded CDF experiment, and show the 
presently achieved detector performances as well as highlights of the RUN II physics 
program in the beauty and electroweak sector.}

\section{Tevatron and RUN II}\label{sec:TeV}
%
%

The Tevatron at the Fermi National Accelerator Laboratory (Fermilab)
with its 980 GeV proton--antiproton beams is currently the highest energy
collider in the world. 
It has resumed operation in April 2001 after a major upgrade. The earlier 
run terminated in 1996 after delivering a luminosity of $\approx~110\rm~pb^{-1}$ 
to both CDF and D0 experiments that allowed a successful physics program ranging 
from the discovery of the top quark to the measurements of W and top masses as 
well as the first evidence for CP violation in B decays. 

The present run (RUN IIa) aims at an integrated luminosity of 
$2 \rm fb^{-1}$ by the end of 2004, a factor 20 more than the entire RUN I. 
To this end the Tevatron accelerator complex 
underwent a substantial upgrade. The Main Ring was dismantled and a new injector 
for Tevatron has been built in a separate tunnel, the Main Injector. It will increase
the production rate of antiproton, a key factor for 
luminosity improvement, by at least a factor three. 
Another major boost in luminosity is expected from the new Recycler Ring,
a permanent magnet storage ring housed in the same tunnel as the Main 
Injector. At the end of a physics store the remaining 75\% of the antiproton beam 
will be decelerated to 8 GeV and stored in the Recycler for later use.
This component of the accelerator complex is still in the commissioning stage 
and will be brought in to operation in 2003. 

In the present configuration Tevatron is running with 36 bunches of protons and 
antiprotons which collide at the interaction regions every 396 ns. 
In a later stage there will be 108 bunches with 132 ns interbunch separation. 
Increasing the number of bunches with respect to the RUN I
configuration allows to keep essentially unchanged the luminosity per
bunch and hence the number of events per crossing thus retaining a
relatively clean environment for offline reconstruction, at the cost of completely
redesigning detector DAQ (see~\ref{subsec:CDFdaq}). 

Fermilab also plans to extend the lifetime of the current 
experiments with a further run (RUN IIb) scheduled to begin in the second half 
of 2005 allowing for a 6 month shutdown for accelerator and detector upgrade.
15 $\rm pb^{-1}$ will be collected by 2008 with the primary focus being the 
search for Higgs Boson in the 115--190 $\rm GeV/c^2$ range.

The first year of the Tevatron operation has been largely devoted to studies on 
the machine and the detectors. As of this writing (May 2002) typical peak 
luminosity of Tevatron is $2~\cdot~10^{31}cm^{-2}s^{-1}$ and CDF is collecting between 
2 and 3 $\rm pb^{-1}$ on tape per week. Tevatron luminosity is expected to increase 
substantially after more cooling power will be installed during summer shutdown.
For the end of 2002 CDF expects to have nearly 100  $\rm pb^{-1}$ on tape 
usable for most of the analysis; presently (May 2002) this figure is 25 $\rm pb^{-1}$.

\begin{figure}
\begin{center}
\psfig{figure=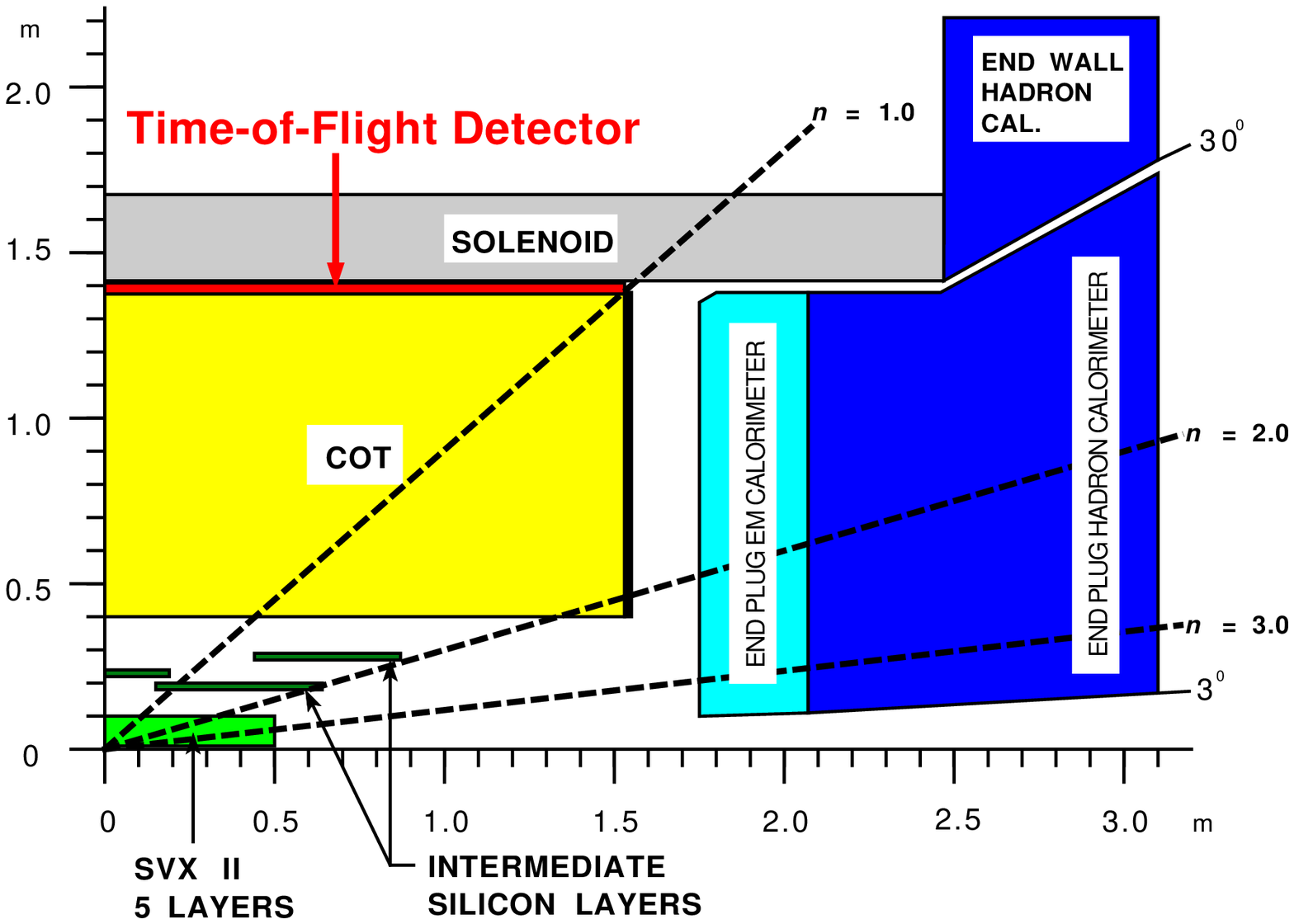,height=1.78in}
\psfig{figure=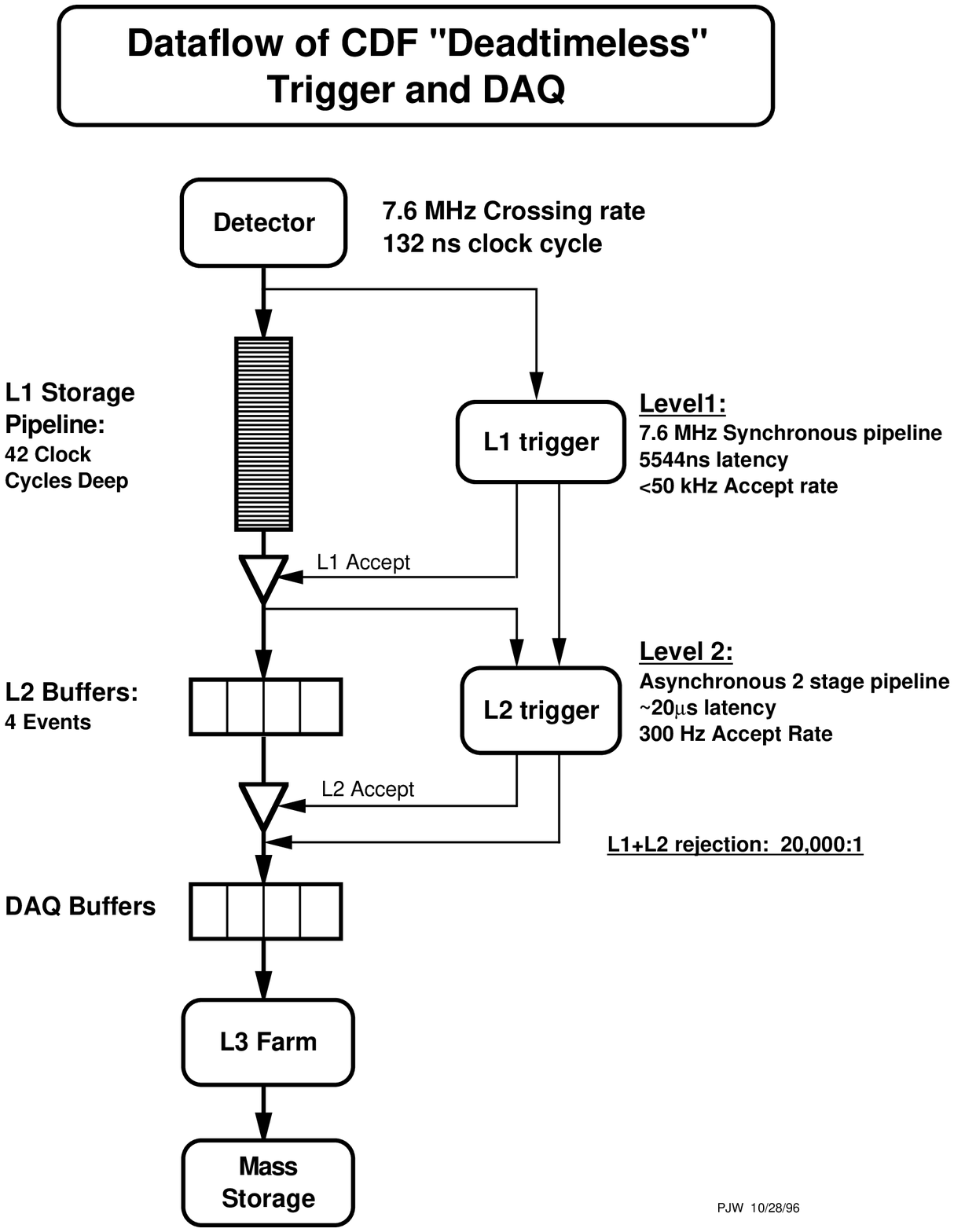,height=1.78in}
\psfig{figure=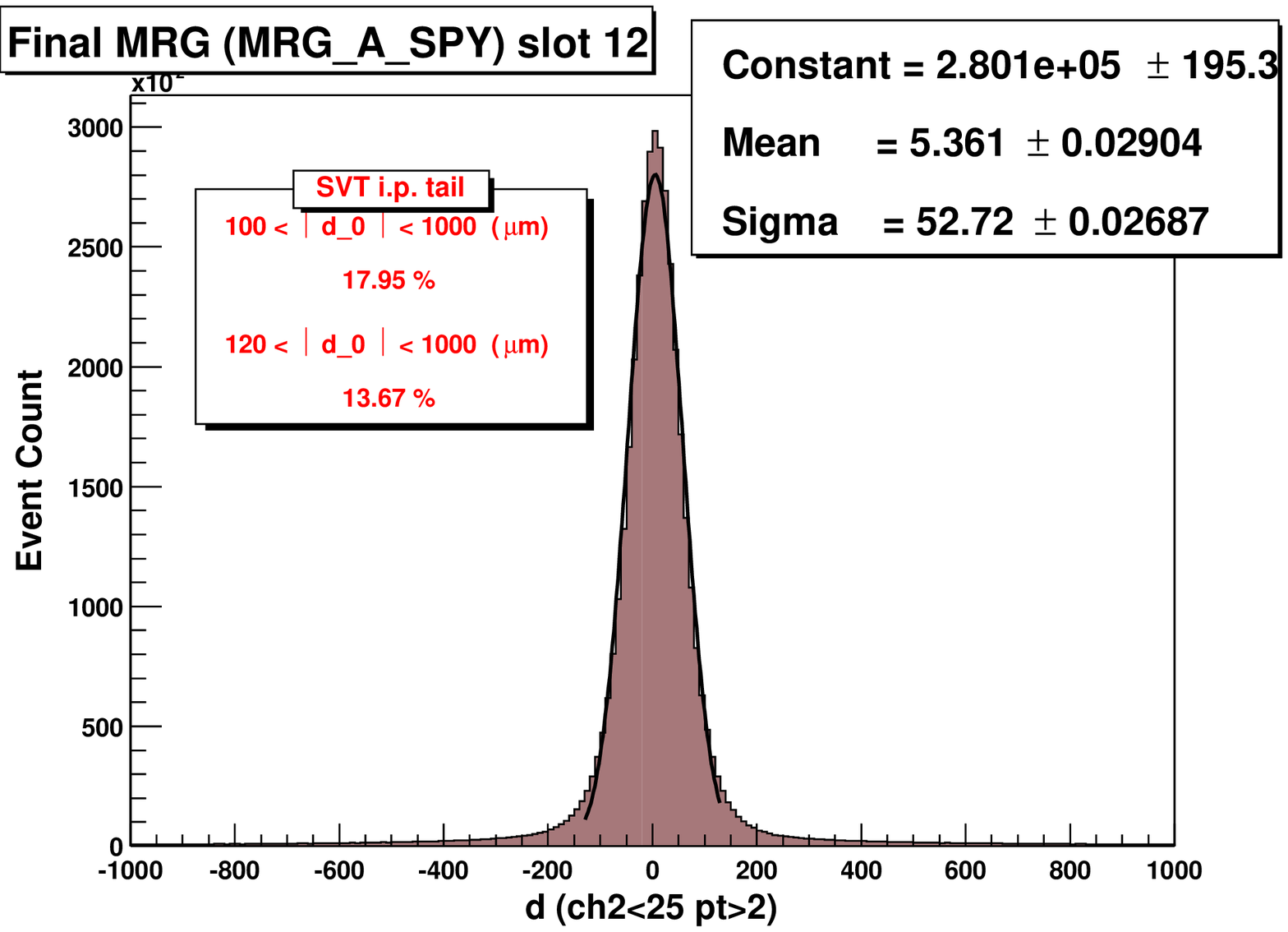,height=1.8in}
\end{center}
\caption{({\it Left}) Quadrant of the CDF detector tracking volume with the plug 
calorimeter;({\it Center}) Trigger and
DAQ flow chart; ;({\it Right}) online SVT impact parameter resolution 
(includes beam width contribution of $\approx 30 \mu m$)}
\label{fig:CDFdaq}
\end{figure}

\section{The CDF detector for RUN II}\label{sec:CDFII}
%
%

The CDF detector  upgrade addressed the issues related to the shorter
bunch crossing time of RUN II with a completely rebuilt DAQ and
Trigger system, new Plug calorimeters, and a faster drift
chamber~\cite{td}. The vertex detector was completely rebuilt with a larger
silicon detector to improve tracking efficiency and enhance b--tagging at
larger rapidity. Triggering at Level 2 on displaced vertex from heavy flavour
decays became possible with the Silicon Vertex Tracker (SVT). Particle
Identification, previously based only on charged particles $dE/dX$ measured in 
the drift chamber, has been complemented with the Time Of Flight (TOF).

\subsection{Trigger and DAQ}\label{subsec:CDFdaq}

All front end electronics has been completely redesigned to cope with
the shorter interbunch separation and the ``deadtimeless'' design of the 
new three stage trigger system. At Level 1 track segments in the muon detector,
calorimeter cell energies and tracks reconstructed in the drift chamber by the
eXtremely Fast Tracker (XFT) processor are used to trigger on tracks,electron,muons,jets
and missing energy. A synchronous pipelined readout allows
a latency for Level1 decision of 5.5\mus\/ with 50 KHz maximum bandwidth
into the Level 2. At this stage 4 alpha processors with a buffered
readout refine the trigger decision using further information from the
SVT processor (see~\ref{subsec:CDFsvt}), the electromagnetic shower max detector, 
the calorimeters and the muon detectors. There is a maximum bandwidth of 
300 Hz towards the L3 trigger which is realized as a farm of
commodities PC running a specialized version of the offline
reconstruction program for the final decision. Events passing L3 requirements are 
written to mass storage with a maximum rate of 75 Hz or $\sim 20$ MB/s.
The DAQ system is performing well and all the triggers foreseen for RUN II have been
implemented by February 2002.

\begin{figure} [t]
\begin{center}
\psfig{figure=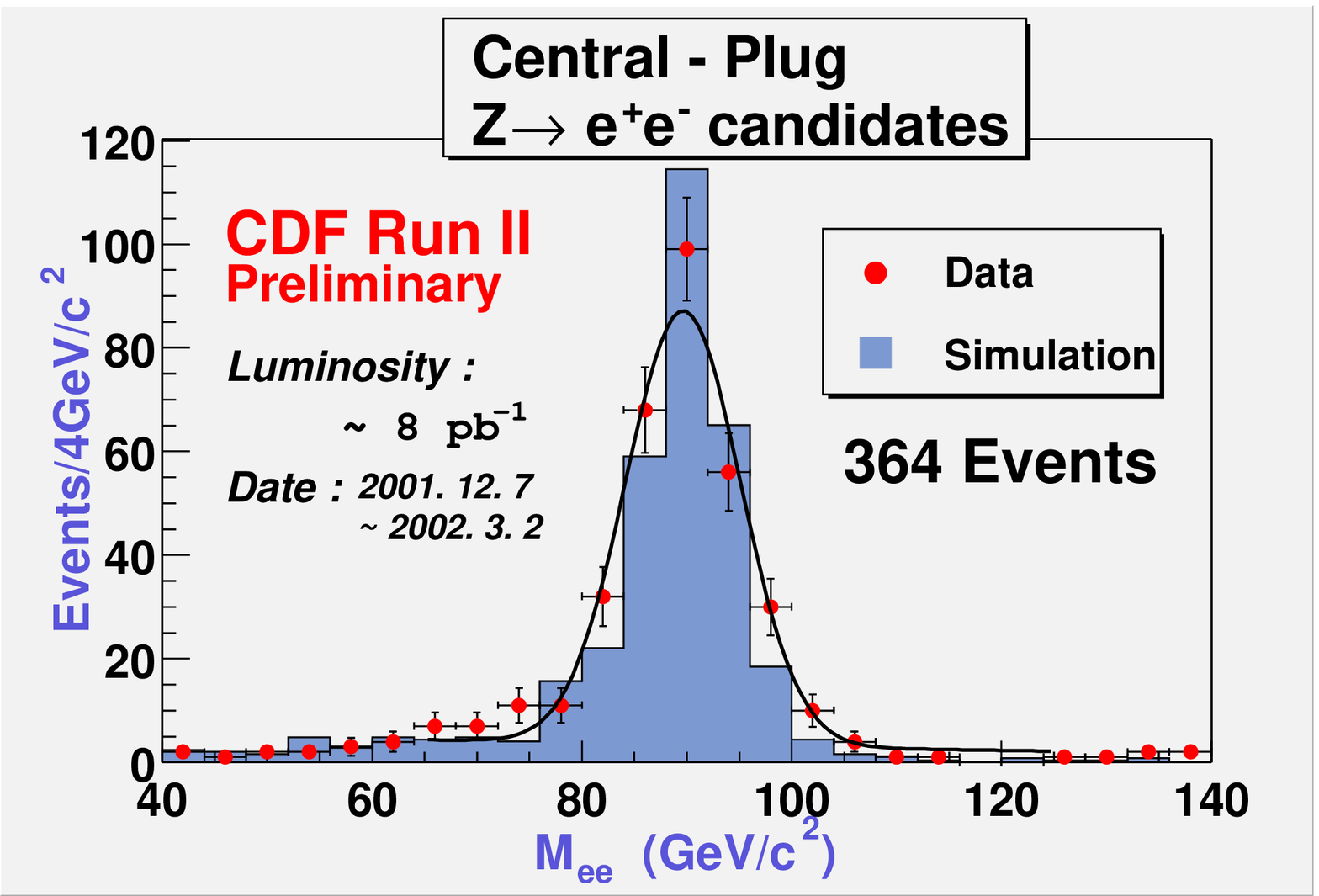,height=2.0in}
\psfig{figure=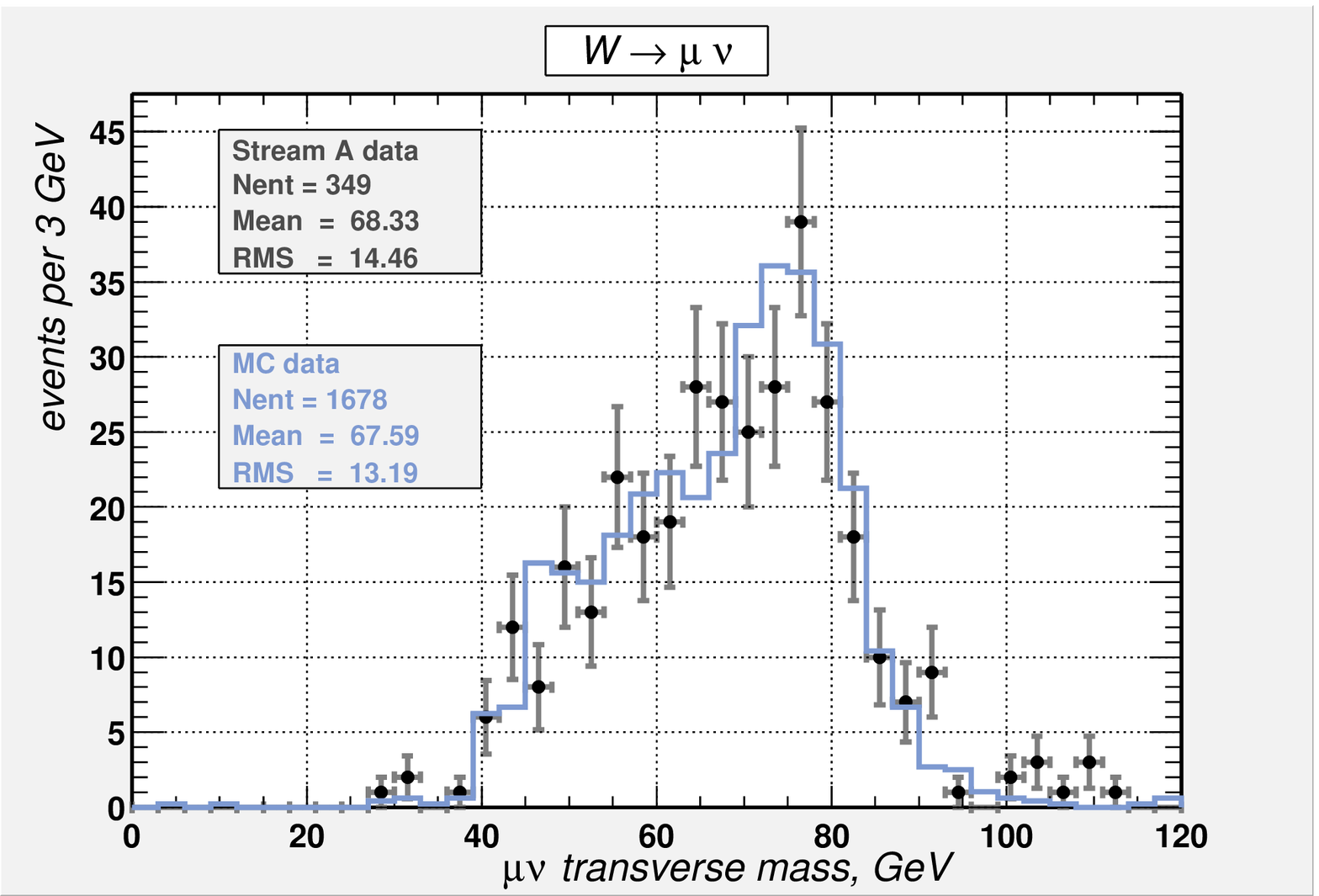,height=2.0in}
\end{center}
\caption{
({\it Left}) $Z \ra e^+e^-$, with $e^+$ or $e^-$ in the new plug;
({\it Right}) Transverse mass for $W\ra\mu\nu$ events
}
\label{fig:highpt}
\end{figure}

\subsection{Tracking}\label{subsec:CDFtracking}

CDF has built a new integrated tracking system combining a large drift
chamber (COT) and three silicon detectors; L00, SVXII and ISL
(fig.~\ref{fig:CDFdaq}).
The COT has a faster drift time with respect to the RUN I detector and a more robust 
design for measuring tracks in the r--z
plane. The combined L00, SVXII and ISL detectors provide 7 high
precision points in the central rapidity region (8 in the high $\eta$ region 
$1<|\eta|<2$). All systems but L00 have double sided sensors.
L00 is the closest to the beam with just 1.6 cm radial distance from
the center of the beam line. The outer ISL layer provides 
measurements at 28 cm radial distance, thus allowing accurate 
track reconstruction using only silicon detectors in the forward region 
$1<|\eta|~<~2$) where the COT acceptance rapidly decreases.
Expected transverse momentum resolution are $\sigma_{p_t}/p_t^2< 0.1 \%$ 
while better than 30 $\mu$m for $p_t>1$ Gev are expected on impact parameter.  
Currently L00 is still under commissioning and is not yet used in the 
reconstruction as is the central part of the ISL system. 
SVXII instead has more than  90\% of its individual silicon 
detectors regularly functioning. 
Alignment and calibration of the detector are currently in progress.
Nevertheless already with a first pass attempt at determining internal alignment,
measured performances confirm expectations. Resolution on impact parameter for 
tracks from prompt $J/\psi$ muons have been measured~\cite{ma} to be 
26 $\mu$m for $p_t>2.2 \rm GeV$.

\subsection{SVT}\label{subsec:CDFsvt}

The SVT~\cite{st} receives silicon detector raw data after each L1
accept. It performs pattern recognition inside the SVXII and L00 systems and
associates silicon hits with tracks from 
the XFT to fit track parameters in the
transverse plane in less than 20 $\mu$s. The resolution on SVT impact
parameter is expected to be similar to that of the offline
reconstructed track thus allowing triggering on displaced vertex from
Heavy Flavour decays. This is the first of such a device ever built
for an hadron collider. The commissioning of the system was quickly
successful and first data triggered by the SVT was taken as early
as October 2001~\cite{sp}. SVT is regularly part of the CDF trigger since
February 2002.  To date, nearly 10 $\rm pb^{-1}$ has been 
integrated with the SVT trigger that currently requires two tracks with opposite
charge and impact parameter greater than 100 $\mu m$. The currently
achieved resolution on impact parameter is on average 52 $\mu m$ including a
beam width contribution of $\approx 30 \mu m$ as shown in fig.~\ref{fig:CDFdaq}. 
This corresponds to 40 $\approx \mu m$\/ resolution to be compared to 
35 $\mu m$\/ design. The level of the tails of the SVT impact parameter distribution, 
which drives the trigger rate, is around 15\%. Both these figures
will improve when better alignment and calibration of the detector
will be available. 


\subsection{Plug calorimeters}\label{subsec:CDFcal}

The endplug calorimeter system has been entirely rebuilt for RUN II,
mainly because the gas detectors of the RUN I calorimeters were not fast enough
for the new timing constraint. Moreover the new detector, built with
scintillating tiles as active medium, has a better sampling fraction and
a better rapidity coverage ($|\eta| <3.6 $) than the old plugs. The
system is operating reliably and calibrations are under way. A clean
signal of $Z\ra e^+e^-$ where one of the electrons is detected in the plug 
calorimeters has been successfully reconstructed and is 
shown in fig.~\ref{fig:highpt}
 
\subsection{TOF}\label{subsec:CDFtof}

\begin{figure} [t]
\begin{center}
\psfig{figure=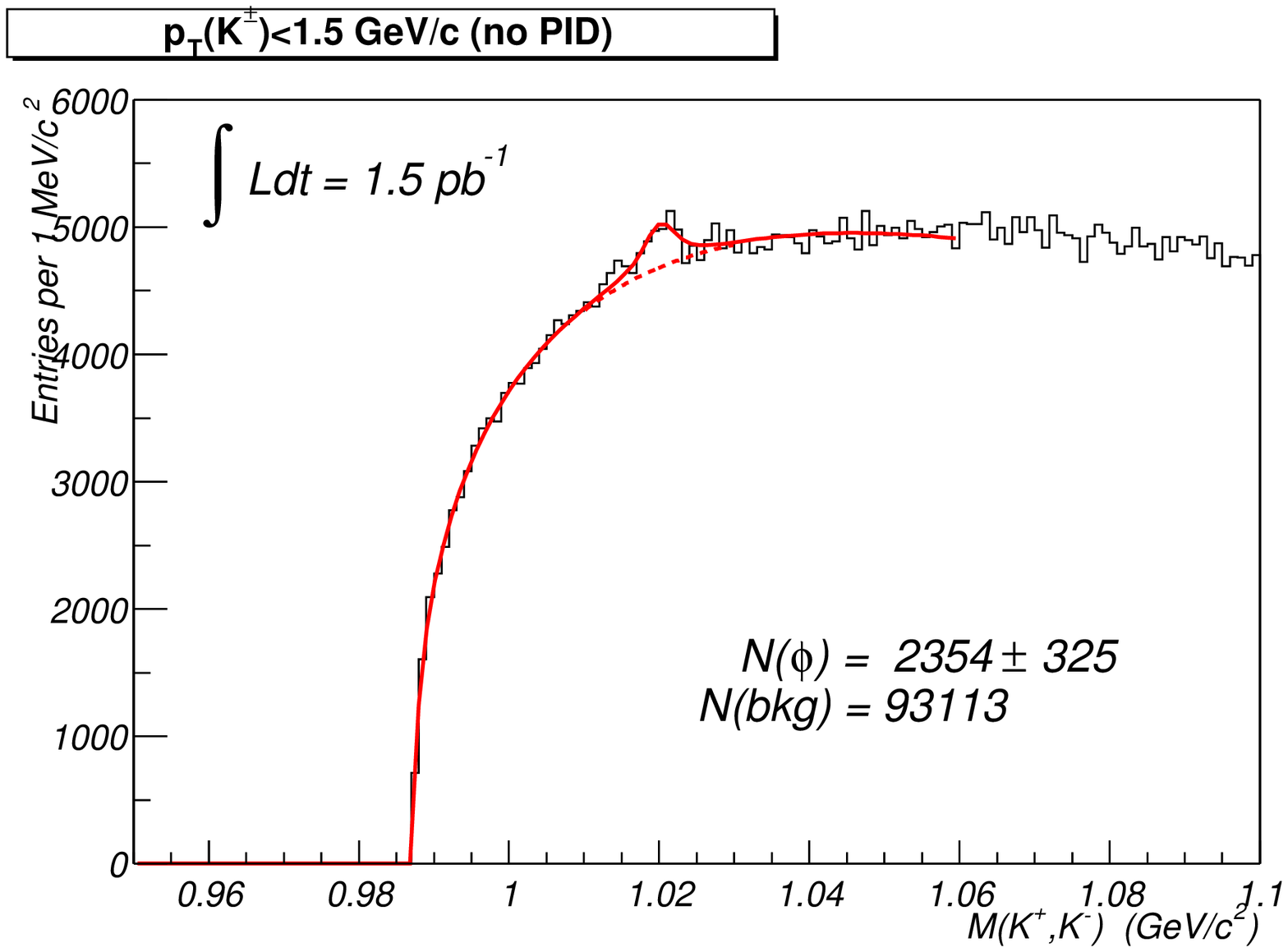,height=1.8in}
\psfig{figure=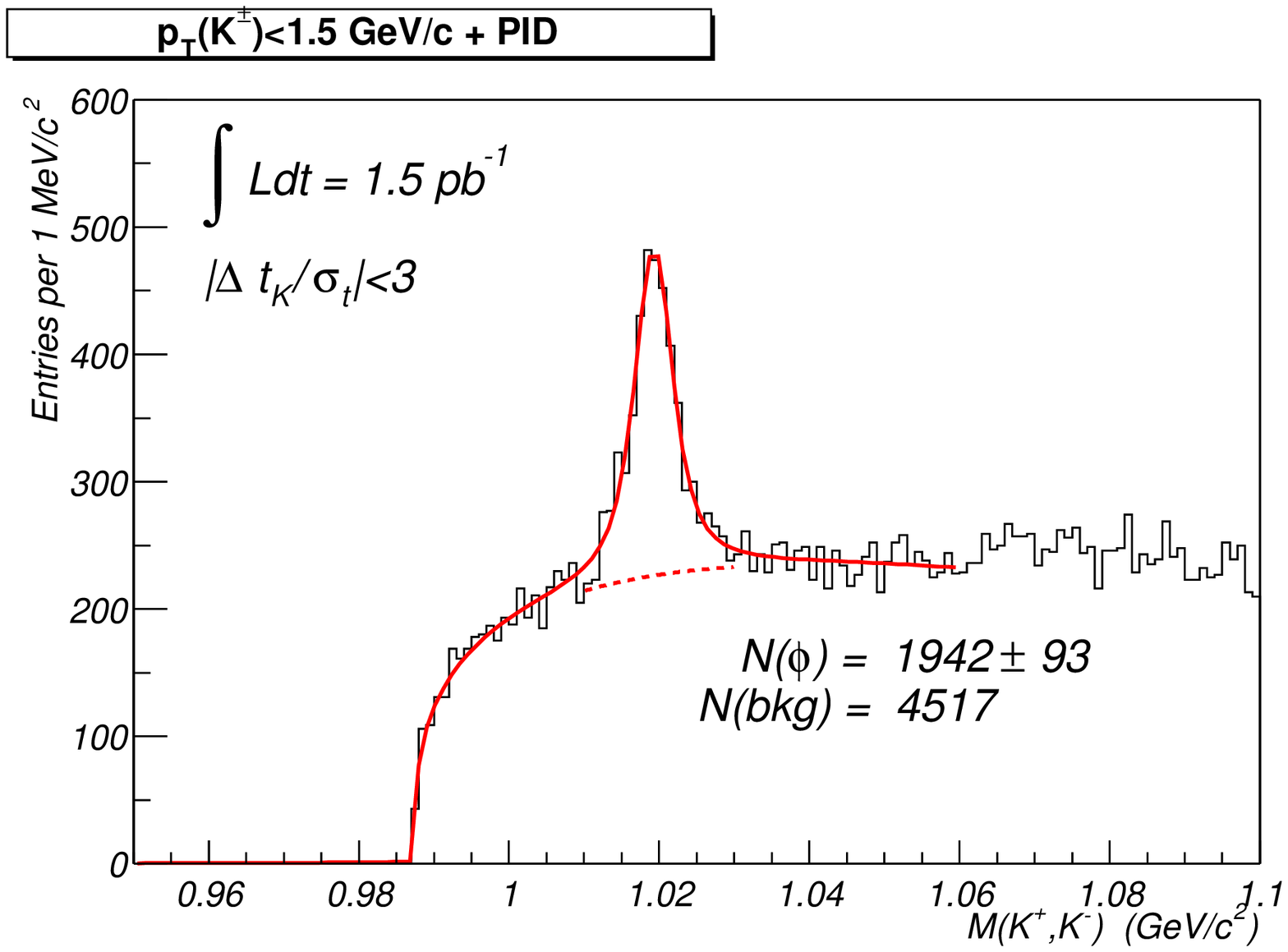,height=1.8in}
\end{center}
\caption{Example of Kaon identification in CDF with the new TOF detector 
to extract a $\phi\rightarrow K^+ K^-$ signal}
\label{fig:tof_phi}
\end{figure}

The Time Of Flight detector is composed of 216 bars of fast plastic
scintillator of approximately 4x4 cm section and 280 cm length, placed in the
space between the COT and the solenoid, and readout by
19 stage Hamamatsu fine mesh phototubes. The design resolution of the TOF
detector is 100 ps that gives a better than $2\sigma$ $\pi/K$ 
separation for $p_t<1.6$ GeV. The TOF detector needs very
careful calibrations to reach the desired performance, that currently are not 
yet finalized. The present measured resolution is 110 ps,
allowing e.g. a clean separation of a $\phi\ra K^+K^-$ signal from 
the combinatorial background as illustrated in fig.~\ref{fig:tof_phi}

\section{First signals from RUN II data}\label{sec:signals}
%
%

Data taken in the first year of operation have been extremely useful
to finish commissioning the detector and trigger. By the start of 2002
CDF has essentially completed this task and is now collecting physics quality
data. Several hundreds W and Z events in both the electron and muon
channels were cleanly reconstructed as early as fall 2001 (fig.~\ref{fig:highpt}),
and were extremely valuable for detector commissioning and calibration.

With the dimuon trigger clean samples of $J/\psi$ and $\psi\prime$ 
were collected. This trigger has a substantially increased acceptance with respect 
to the RUN I configuration, and the expected increase in event yield has been observed 
in RUN II data. The width of the peaks are consistent with expectation too, and 
a preliminary attempt at measuring inclusive b lifetime using $b\ra J/\psi X$ 
events gives already results consistent with world average making us confident
on the quality of the detector and of its present alignment.

\begin{figure}
\begin{center}
\psfig{figure=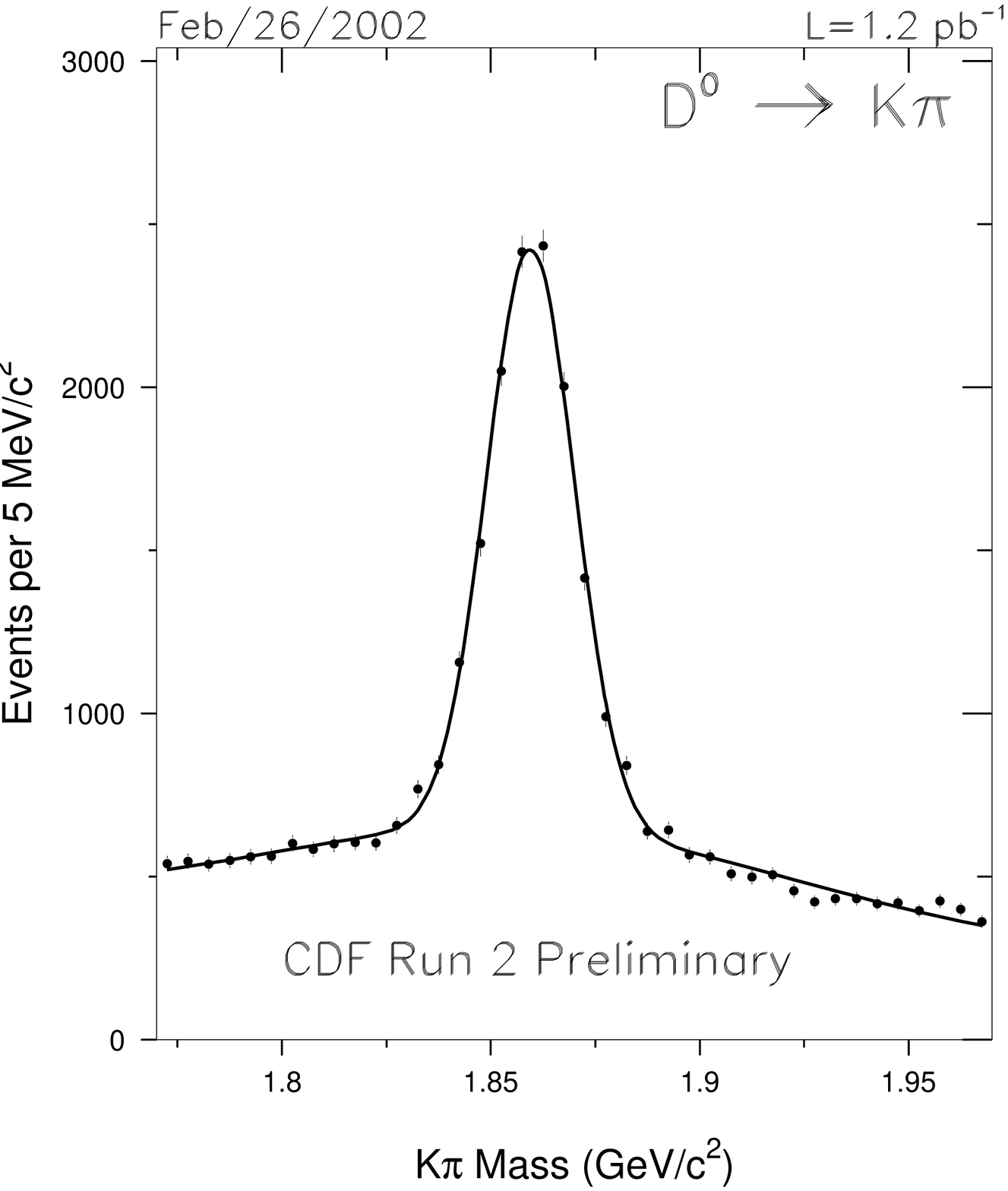,height=1.8in}
\psfig{figure=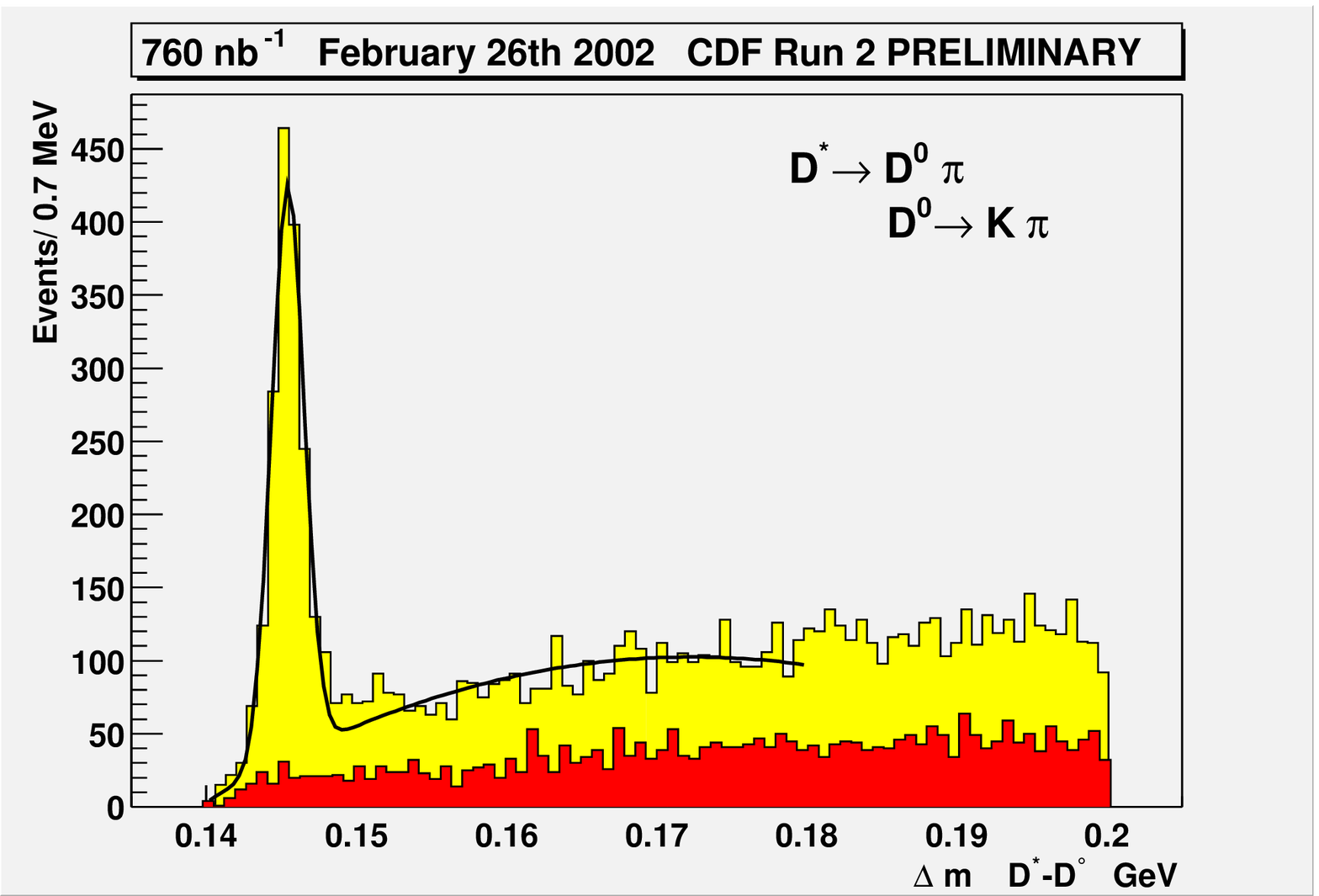,width=2.5in}
\psfig{figure=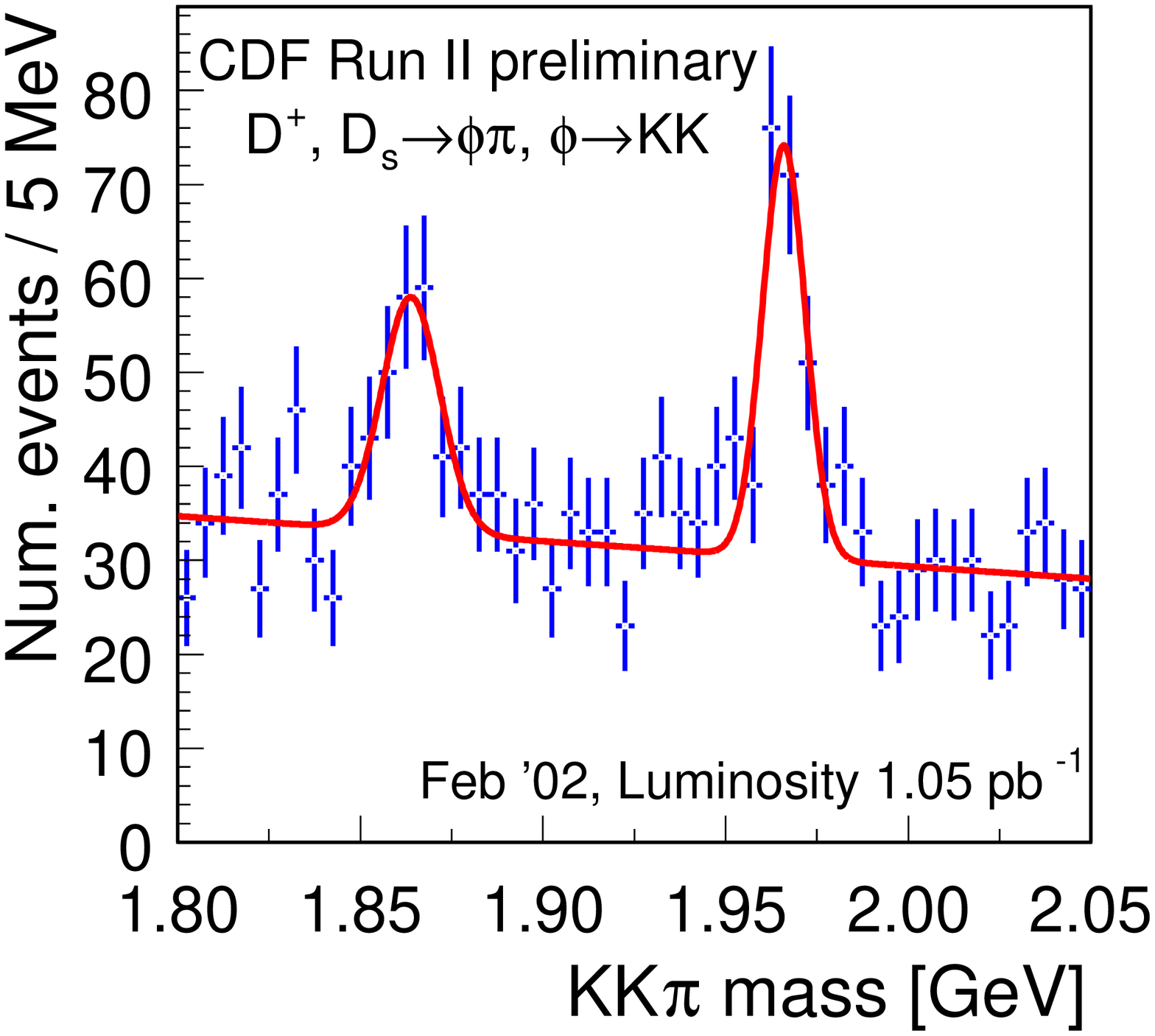,height=1.8in}
\end{center}
\caption{Charm meson peaks in the displaced vertex trigger. From left to right: 
$M_{k\pi}$ for $\Do \ra K\pi$; $\Delta M$ for $D^{*+}\ra \Do \pi^+$; $M_{kk\pi}$ for
$D_s\ra \phi \pi^+ \ra K^+K^-\pi^+$  and the identical Cabibbo suppressed $D^+$ decay}
\label{fig:charm}
\end{figure}

The ability to trigger on displaced vertex through the SVT is the most important 
novelty for b--physics in CDF. It was thus very important that clean 
signals of charm mesons were quickly established in the first data 
(less than 1 $\rm pb^{-1}$) collected with this trigger. 
In fig.~\ref{fig:charm} peaks from the decays $\Do \ra K\pi$, 
$D^{*+}\ra \Do \pi^+$ and $D_s\ra \phi \pi^+ \ra K^+K^-\pi^+$ are shown. 
No particle identification is yet used in the analysis as well as the 3D 
capability of the silicon detectors, signal extraction thus rely only on mass 
separation. Nevertheless very encouraging signal to noise ratio were obtained 
just repeating offline the trigger selections, as an example a $S/B > 3$ was 
obtained on the $\Do \ra\ K\pi$ peak . 
This was an extremely positive 
finding suggesting that despite hostile hadronic environment relatively 
clean sample of b hadrons will eventually be reconstructed in the SVT triggers,
and that it will be possible to keep the QCD background at a manageable level. 
Moreover given the measured 
yields of charm mesons (e.g. $\approx 8$\/ nb for $D^0$) CDF is soon 
expected to collect samples bigger than those presently collected at the B factories. 
These data will be used for unique measurements of charm production in 
$p\bar{p}$ collisions and may possibly lead to competitive measurements of 
charm mixing and CP violation parameters.

\section{Physics prospects in RUN IIa}\label{sec:highlights}
%
%

\subsection{B physics}\label{subsec:Bphysics}

Tevatron and the CDF detector offer unique opportunity for studying all species of 
b hadrons taking advantage of the huge cross section for b production in 
$p\bar{p}$ collision at $\sqrt{s}=2$ TeV. Beside the spectroscopy 
and lifetime measurement for a large number of b hadrons, including e.g. $B_c$,
CDF can put significant new constraints on the CKM parameter space.

Estimates~\cite{yb} based on extrapolation of RUN~I data and expected performances 
of the new trigger and detectors indicate that a very competitive measurement of 
$sin(2\beta)$ through the observation of the time dependent asymmetry of 
$B^0$ and $\bar{B^0}$ in CP states will be possible with RUN~IIa luminosity. 
The expected error with 2fb$^{-1}$ of integrated luminosity is 0.05 using only 
the ``golden'' mode $B^0(\bar{B^0}) \ra J/\psi K^0_s$ with the 
$J/\psi \ra \mu^+\mu^-$.
The key factors in this measurement are the accumulation of large samples of 
these decays and the effective power of the experiment in tagging the initial 
flavour of the B meson.
This estimate is based on the expected 20000 $B^0(\bar{B^0})\ra J/\psi K^0_s$ decays
that will be collected by CDF in RUN II taking into account the increased 
acceptance of the muon detectors and the lower $p_t$ thresholds of the new trigger. 
With the RUN II detector CDF expect a combined $\epsilon D^2$ (the product of the 
tagging efficiency $\epsilon$ and the dilution factor $D=1-2w$, where $w$ is the wrong 
tagging probability) of 9.1\%~\cite{yb}. The factor 2 increase over 
the RUN I analysis~\cite{2b} is due in part to the extended acceptance for tracks 
of the new integrated tracking, which improve the performance of the Jet--charge 
tag, and in part to the introduction of the Opposite Side Kaon tag 
with the new TOF detector~\cite{bb}.
 
Of particular importance in constraining the CKM matrix is the measurement 
of the mixing parameter $x_s=\Delta m_s/\Gamma$ in $B_s$ flavour oscillation. 
The Standard Model fit to world data prefer for $x_s$ the range 
$22.3 < x_s < 31.3$, while the present combined world lower limit 
on $\Delta m_s>14.9 \rm ps^{-1}$ @ 95\% C.L~\cite{m1}. In addition to the usual 
semileptonic mode, CDF will collect large 
samples of completely reconstructed hadronic decays of $B_s$ with the SVT trigger 
that will be extremely valuable for the oscillation measurement especially if the 
$x_s$ parameter turn out to be greater than 30. In fact the superior proper time 
resolution achievable in completely reconstructed modes, combined with 
the excellent impact parameter resolution of the L00 silicon detector, will 
allow a $5\sigma$ measurement with RUN II data up to $x_s \approx 60$~\cite{yb}.
As in the $sin(2\beta)$ analysis also in the the $B_s$ mixing the new TOF 
detector plays a key role, greatly improving the effective tagging efficiency 
of the experiment, with the Same Side Kaon tag which correlates the charge of the 
Kaon produced in association with $B_s(\bar{B_s})$ in the hadronization process 
with its initial flavour~\cite{bb}.

\subsection{High $P_t$ physics}\label{subsec:highpt}

RUN II has the potential for very interesting results in the Electro--Weak and 
top sector.

The increase in luminosity and lepton acceptance will lead
to statistical error of 20 Mev in the W mass measurement for both the electron and 
muon channel~\cite{sm}. The most important source of systematics in the 
RUN I result~\cite{mw} was the lepton energy scale and resolution, as 
determined by $Z\ra l^+l^-$  data, and was finally limited by the available 
statistic. This error is expected to scale with luminosity and be $\sim 13$~MeV
with RUN IIa data. The leading error will then come from uncertainties in 
W production and decay models, also in part constrained by collider data, 
that will be $\sim 15$ MeV. A measurement for a single channel and single experiment 
with 40 MeV error seems feasible and is well matched to the present LEP2 
precision as well as the current uncertainty in the indirect determination of $M_W$ 
from Standard Model fit~\cite{wl}. For the W width an error of $\sim 50$ MeV is foreseen.

The top mass uncertainty will be greatly reduced in RUNII, with a statistical error 
of 1.7 GeV in the lepton + 2 b--tag sample alone. This is the channel 
with the greatest sensitivity to the top mass and will benefits from the  
improved b--tagging efficiency expected with RUN II detector. Jet energy scale
uncertainty may be reduced using both $Z\ra b\bar{b}$ events and reconstructing the 
W mass from jets in lepton + 2 b--tag top sample. The goal for RUN IIa is a 
combined measurement with 2 GeV error.

\section{Conclusions}

The CDF detector has essentially completed its commissioning phase by the end of
year 2001 and has begun to collect data for RUN II. The detector performs as 
expected and first results will appear for Summer conferences in 2002.


\section*{References}
\begin{thebibliography}{99}

\bibitem{td} R.Blair (the CDF Collaboration), FERMILAB-PUB-96/390-E(1996),
{\em The CDF II Detector Technical Design Report}.

\bibitem{ma}M.Bishai, private communication.

\bibitem{st}
A.~Bardi {\it et al.},
{\em SVT: An online silicon vertex tracker for the CDF upgrade,}
Nucl.\ Instrum.\ Meth.\ A {\bf 409} (1998) 658.

\bibitem{sp}
W.~Ashmanskas {\it et al.},
{\em Performance of the CDF online silicon vertex tracker,}
FERMILAB-CONF-02-035-E
{\it Presented at 2001 IEEE Nuclear Science Symposium (NSS) and Medical Imaging Conference (MIC), San Diego, California, 4-10 Nov 2001}.

\bibitem{yb}
K.~Anikeev {\it et al.},
{\em B physics at the Tevatron: Run II and beyond,}
arXiv:hep-ph/0201071.

\bibitem{2b}
T.~Affolder {\it et al.}  [CDF Collaboration],
Phys.\ Rev.\ D {\bf 61} (2000) 072005

\bibitem{bb}
``Proposal for Enhancement of the CDF II Detector: ...'' (P-909);
''Update to Proposal P-909:Physics Performance ...''
Documents available at http://www-cdf.fnal.gov/upgrades/btb.

\bibitem{m1}
A.Sciab\`a, this proceedings.

\bibitem{mw}
T.~Affolder {\it et al.}  [CDF Collaboration],
Phys.\ Rev.\ D {\bf 64} (2001) 052001

\bibitem{sm}
M.~Grunewald, U.~Heintz, M.~Narain and M.~Schmitt,
arXiv:hep-ph/0111217.

\bibitem{wl}
C.Parkes, this proceedings; G. Myatt, this proceedings.

\end{thebibliography}
\end{document}